\documentclass{aa}
\usepackage{graphicx}
\usepackage{supertabular}
\begin{document}
   \title{Southern Infrared Proper Motion Survey III: Constraining the mass function of low mass stars} 
   \titlerunning{Southern Infrared Proper Motion Survey III}
   \subtitle{}

   \author{N.R. Deacon
          \inst{1}, G. Nelemans \inst{1} \and N. C. Hambly\inst{2}   
          }
   \authorrunning{Deacon, Nelemans \& Hambly}
   \institute{\inst{1}Department of Astrophysics, Faculty of Science, Radboud University Nijmegen, P.O. Box 9010, 6500 GL Nijmegen, The Netherlands\\
  \inst{2}SUPA\thanks{Scottish Universities' Physics Alliance}, Institute for Astronomy, University of
              Edinburgh, Royal Observatory, Blackford Hill, Edinburgh EH9 3HJ, UK\\
              \email{ndeacon at astro.ru.nl}
             }

   \date{Received ---; accepted ---}

   \abstract{ The stellar mass function is one of the fundamental distributions of stellar astrophysics. Its form at masses similar to the Sun was found by Salpeter (1955) to be a power-law $m^{-\alpha}$ with a slope of $\alpha=1.35$. Since then the mass function in the field, in stellar clusters and in other galaxies has been studied to identify variation due to environment and mass range. Here we use results from previous papers in the SIPS series to constrain the mass function of low mass stars (0.075M$_\odot$$<$m$<$0.2$_\odot$). We use simulations of the low mass local stellar population based on those in Deacon \& Hambly (2006) to model the results of the SIPS-II survey (Deacon \& Hambly, 2007). We then vary the input parameters of these simulations (the exponent of the mass function $\alpha$ and a stellar birthrate parameter $\beta$) and compare the simulated survey results with those from the actual survey. After a correction for binarity and taking into account potential errors in our model we find that $\alpha=-0.62\pm0.26$ for the quoted mass range.
   }

   \maketitle
%

\section{Introduction}
The mass function was defined by Salpeter (1955) to be $\xi(\log_{10}(m))=\frac{dn}{d\log_{10}(m)}$. Its form and variation are still hot topics in astrophysics. The original Salpeter (1955) paper built on the work of Luyten (1941) and Van Rhijn (1936) to produce a luminosity function. This was converted into a mass function fitted by a power law ($\xi(\log_{10}(m))=m^{-\alpha}$) with an exponent of 1.35. This was applicable over a range from 0.4 to 10 solar masses. Since the late 1970s studies have begun to identify a a shallowing and even a turnover in the mass function at lower masses (see Miller \& Scalo (1979)). Reid et al. (2002) used Hipparcos data and their own discoveries to produce a volume limited sample down to $M_V = 15.5$. Using this V band luminosity function they fitted a power law with the exponent $\alpha = 1.35 \pm 0.2$ in the region $0.1 $M$_\odot < m < 1.0 $M$_\odot$. They also produced a volume limited 8pc sample with a value of $\alpha$ of $1.15 \pm 0.2$ in the same region. Kroupa (2001) took the luminosity function approach as well as examining underlying problems such as binarity. He fitted a four segment power law such that for $0.01 $M$_\odot < m < 0.08 $M$_\odot$ being fitted by
$\alpha = -0.7\pm0.7$, $0.08 $M$_\odot < m < 0.5 $M$_\odot$ by $\alpha =
0.3\pm0.5$, $0.5 $M$_\odot < m < 1.0 $M$_\odot$ by $\alpha = 1.7\pm0.3$
and $1.0 $M$_\odot < m$ by $\alpha = 1.3\pm0.7$. Allen et al. (2005) used a
 series of assumptions about the birthrate and a Bayesian method to
 yield a value of $-0.7$ in the range $0.04 $M$_\odot < m < 0.1
 $M$_\odot$. Chabrier (2001) used
both $V$ and $K$ band volume limited 5pc Luminosity Functions to produce a
Mass Function well fitted by a lognormal form (a parabola in log-log space) peaking at 0.08 solar masses. This work was superceded by Chabrier (2005) where a re-evaluation of the local Luminosity Function indicated a mass function peaking at 0.2M$_\odot$. Zheng et al. (2001) use HST data along with a model of metallicity vs. scale height to estimate $\alpha = -0.1$ in the region 0.5 - 0.1 solar masses with no binary correction, becoming $\alpha = -0.45$ after binary sorrection. Tinney (1993) reports that the mass function peaks and declines below 0.2 M$_{\odot}$. However he also reports a rise towards the hydrogen burning limit, something not reported in recent studies. Martini \& Osmer (1998) fit a virtually flat ($\alpha = 0.32 \pm 0.15$) mass function between 0.6 and 0.1 solar masses. Uniquely among recent studies Schultheis et al.'s study using CHFT data gives a mass function below 0.2M$_{\odot}$ that has a slope more steeply increasing ($\alpha = 2.0$) than Salpeter's.

Inferrring the form of the mass function from the luminosity function is relatively easy in single age populations such as open clusters. This is simply a conversion using a mass-luminosity relation. However when dealing with the field population - where the ages of stars are spread over the whole history of the Galaxy - this becomes more difficult. Hence we must also consider the birthrate $b(t)=\frac{dn}{dt}$. Schmidt (1959) predicted a declining birthrate based on the assumption that it is related to the density of interstellar gas.
However, as noted by Miller \& Scalo (1979), because his work came before
the widespread acceptance of a hot Big Bang (and hence primordial
nucleosynthesis) he uses an initial helium abundance of zero. Clearly
this is not correct.  Miller \& Scalo (1979) went on to use a continuity
constraint, that the mass function should be smooth to derive the
birthrate. They rejected Schmidt's declining form instead
preferring a roughly constant birthrate. Some studies claim the birthrate is not smoothly varying, for example Rocha-Pinto et al. (2000) used the chromospheric activity of stars to derive their ages. From this age distribution they yield a birthrate that
shows a series of three or four bursts of star formation in the last
10Gyrs. They note some of the
bursts may be linked to close encounters with the Large Magellenic Cloud.

Finally any study using only surveys must include a correction for unresolved binarity (which will both alter the colours of objects and hide cool, dim companions). The binary fraction of higher mass stars is fairly high. G dwarfs for example have a binary frequency of 57 $\pm$ 7\% (Duquennoy \& Mayor, 1991). For low mass stars the fraction may be lower. Fisher \& Marcy (1992) found 42 $\pm$ 9\% for M0-M4 dwarfs and Maxted \& Jeffries (2005) founnd a binary fraction of 32-45\% for stars below about 0.15M$_\odot$. 

In this paper we model the SIPS-II survey (Deacon \& Hambly, 2007). This is a proper motion survey combining SuperCOSMOS scans of UKST $I$ plates (Hambly et al., 2001) and data from the 2MASS survey (Skrutskie et al., 2006). This produced a sample of roughly 7000 low mass stars with proper motions between a tenth and half an arcsecond per year. We will outline how we modelled this survey and how these models can be used to constrain the underlying mass function and birthrate.

\section{Simulation Method}

The method used here is based on the simulations used for Deacon \& Hambly (2006). These are similar in principle to those of Burgasser (2004). A short overview of the techniques used in Deacon \& Hambly (2006) follows. Individual objects are assigned masses and ages drawn from a mass function (with masses between 0.5 and 0.03 solar masses) and a birthrate (with ages up to 10Gyr). These are then used to derive the apparent magnitudes, space positions and space velocities of each object. These can be used to calculate observable parameters such as proper motion, sky position and apparent magnitude. These parameters are then passed through a survey selection mechanism to yield the results of the simulated survey. By varying the input parameters of the simulated survey the results will also vary. These can then be compared with the actual results to constrain those input parameters. Below we outline the calibrations to produce the photometric and astrometric parameters of each simulated object and the selection mechanism imposed on them. Additionally how the input parameters are constrained is also outlined. 

\subsection{Photometric Simulation}
\begin{figure}[htb]     
        \begin{center}
\resizebox{\hsize}{!}{\includegraphics[scale=1.0]{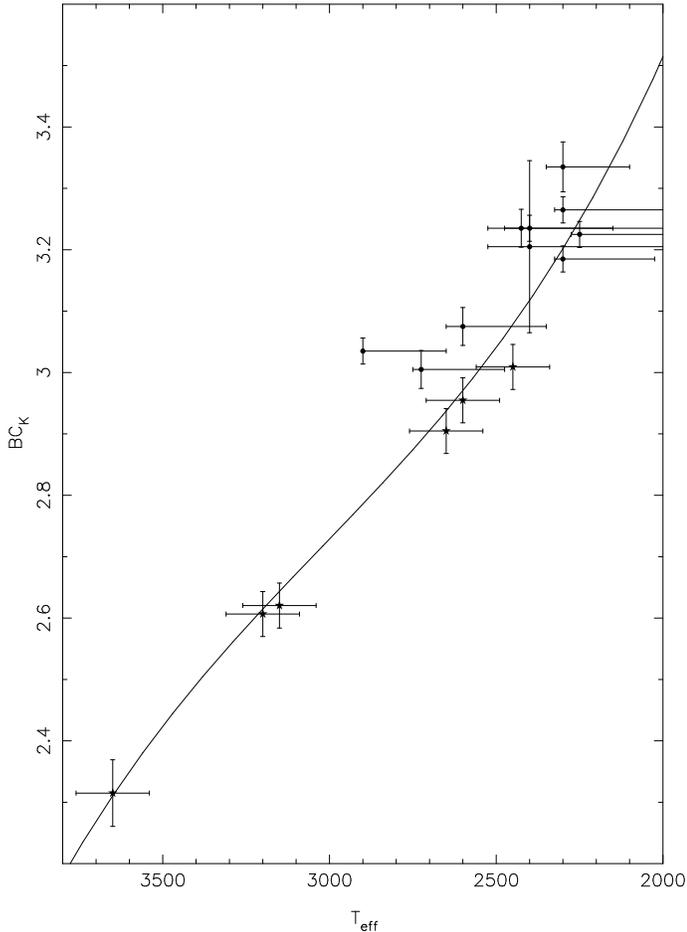}}
\end{center}
\caption{The $K_s$ band bolometric correction vs effective temperature relation. The dots represent data points from Golimowski et al. (2004) and the stars symbols plot data from Berriman \& Reid (1987). Halo objects and objects in binaries have been excluded from the sample. The line is a polynomial fit to the data. Note the scatter on these data points, this is probably due to measurement errors both in the determination of apparent magnitudes but more importantly to the errors in the trigonometric parallaxes of the objects. In the Baraffe et al. (1998) models a 0.5 solar mass star would have a temperature of 3650K and a 0.1 solar mass star a temperature of around 2900K.}
         \label{Tfit}
\end{figure}
\begin{figure}[htb]     
        \begin{center}
\resizebox{\hsize}{!}{\includegraphics[scale=1.0]{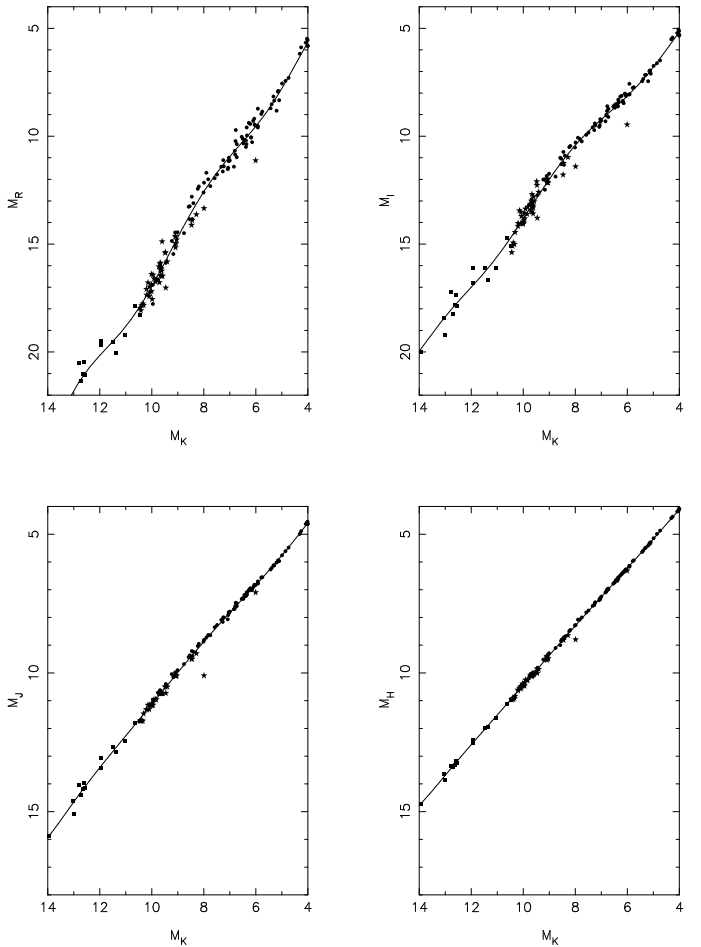}}
\end{center}
\caption{The relation between absolute $K_s$ magnitude and the absolute magnitudes in other passbands. The dots are taken from Reid et al (2002)'s 8pc sample, the squares are from Dahn et al (2002) and the stars are a subset of Cruz et al (2007)'s sample. In each case there is a polynomial fit to the data plotted. Note the larger scatter (much of it intrinsic) on both the $R$ and the $I$ data.}
         \label{Tfit1}
\end{figure}
\begin{figure}[htb]     
        \begin{center}
\resizebox{\hsize}{!}{\includegraphics[scale=1.0]{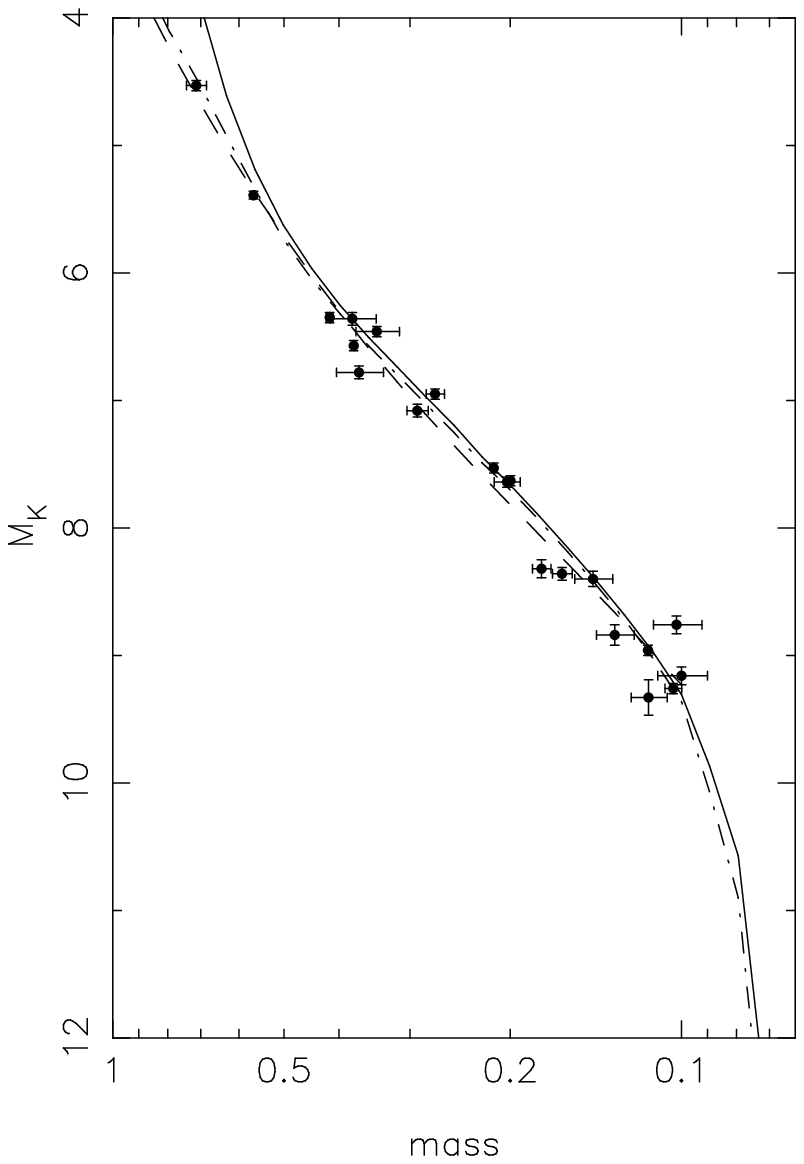}}
\end{center}
\caption{A plot showing the points of binary data from Delfosse et al. (2000). Along with these is our model for a 5Gyr old population (solid line), a fit to the Delfosse data points (dashed line) and the raw Baraffe (1998) and Chabrier (2003) models (dot-dash line) also for a 5Gyr old population. There is no glaring discrepancy. Note the fit to the Delfosse points is not extrapolated below 0.1M$_{\odot}$ as there is no data below this point}
         \label{modelplot}
\end{figure}
Once the masses and ages for a set of objects have been assigned their photometric characteristics can be calculated. Effective temperatures and bolometric magnitudes for these objects are found from evolutionary models (from Baraffe et al., 2003 for objects below $0.1M_\odot$ and from Baraffe et al., 1998 for objects in the range $0.1M_\odot < M < 0.5M_\odot$). In order to ground our simulations in observational data as well as theoretical models an empirical photometric model was used. An effective temperature vs $K_s$ bolometric correction relation was derived from the data contained in Golimowski et al. (2004) and Berriman \& Reid (1987) with the photometry converted to the 2MASS system using the conversions in Carpenter (2001)~\footnote{Note here we ignore the T dwarf population as none appear in the SIPS-II sample.}. The polynomial fit for the effective temperature - $K_s$ bolometric correction relation is shown in Figure~\ref{Tfit}. The $K_s$ absolute magnitudes could then be calculated from the objects' bolometric magnitudes. To test that this method of obtaining $K_s$ was not flawed we plotted $M_K$ vs. mass for our model at 5Gyrs and the models of Baraffe (1998) and Chabrier (2003) against the empirically measured masses and $M_K$s from Delfosse et al. (2000). Clearly there is no significant offset. We had decided against using the Delfosse data in our models as it would not allow us to take into account the luminosity evolution of objects. These $K_s$ band absolute magnitudes are then converted to $R$, $I$, $J$ and $H$ magnitudes using a relation between these magnitudes and the $K_s$ magnitude. This was found by taking data from Reid et al. (2002)'s eight parsec sample, Cruz et al. (2007) and Dahn et al. (2002) and fitting polynomials to these data. The $K_s$ to other passband fits are shown in Figure~\ref{Tfit1} and their coefficients are given in Table~\ref{fits}. Finally each object was given a small offset in $R$ (0.25 magnitudes) and $I$ (0.12 magnitudes) magnitude to simulate the scatter in these passbands on the HR diagram. These values were calculated by allowing the offsets to vary as free parameters and selecting those that gave the best fit. The offset values are also comparable with the scatter around our $R$ and $I$ to $K_s$ relations once measurement errors in the sample used to produce the photometric model are taken into account. By doing this we can take into account the effect the intrinsic scatter has on our magnitude limited sample. However we do not model a scatter that is metallicity dependent and hence we may miss some age/metallicity dependence.
\begin{table*}
   \caption[]{The polynomial fit used in the model. The fits take the form $M_X = \sum_i a_i M_K^i$. As in the SIPSII survey the $R$ is $R_59$, the $I$ band $I_N$ and the $J$, $H$, and $K_s$ bands are on the 2MASS system. Objects in the fit in different photometric systems had their magnitudes converted using Bessell (1986) for the optical data and Carpenter (2001) for the infrared.}
         \label{fits}
\begin{tabular}{lcccccccc}
\hline
&$a_0$&$a_1$&$a_2$&$a_3$&$a_4$&$a_5$&$a_6$&RMS error\\
\hline
$M_R$&1.063e+01&-1.265e+01&6.343e+00&-1.333e+00&1.438e-01&-7.656e-03&1.591e-04&0.433\\
$M_I$&8.075e+00&-9.450e+00&4.890e+00&-1.031e+00&1.102e-01&-5.768e-03&1.175e-04&0.315\\
$M_J$&1.064e+00&-5.835e-01&8.581e-01&-1.886e-01&2.033e-02&-1.063e-03&2.160e-05&0.076\\
$M_H$&1.202e-01&8.509e-01&7.639e-02&-1.412e-02&1.294e-03&-5.493e-05&8.356e-07&0.039\\
\hline
\end{tabular}
\end{table*}
\subsection{Astrometric Simulation}

There are two distinct parts to the astrometric simulation, the velocity simulation and the space positions simulations. The velocity simulations use a simple thin disk model (we see no prominent seperate thick disk population in our sample). These had velocity dispersions $(\sigma_U,\sigma_V,\sigma_W)=(32.6,20.0,15.1)$ (Seabroke \& Gilmore, 2007) which are typical values for a thin disk model. These are assigned randomly and are not related to space positions. A solar reflex velocity was also added with values $(U,V,W)=(10.5,18.5,7.3)$ (Makarov \& Murphy, 2007). The positons are also based on a simple thin disk model with a scale height of 300pc, in the small region of the Galaxy our survey covers (d$<$300pc) we neglect the scale length of the disk which is typically 3500pc (de Vaucouleurs \& Pence, 1978). Hence we assign random $x$ and $y$ positions in the Galactic plane drawn from a flat distributions and a $z$ position out of the plane drawn from a declining exponential with a scale height of 300pc~\footnote{Tests using an age dependent Galactic model did not produce substantially different results.}.    

\subsection{Selection Mechanism}
Once the basic physical properties of the simulated star are calculated it can then be passed through a survey selection mechanism. Before this can be done the observational characteristics of each object (sky position, apparent magnitude, proper motion) are calculated. Also errors are added to simulate the real survey. These add an error to the photometry and an error to the position which will affect the proper motion. Each of these errors is proportional to the stars brightness and is derived from the errors quoted for both the UKST and 2MASS surveys (Hambly et al., 2001 and Skrutskie et al., 2006), at best 0.1 magnitudes in the UKST data and 0.02 in 2MASS.

\begin{figure}[htb]     
        \begin{center}
\resizebox{\hsize}{!}{\includegraphics[scale=1.0]{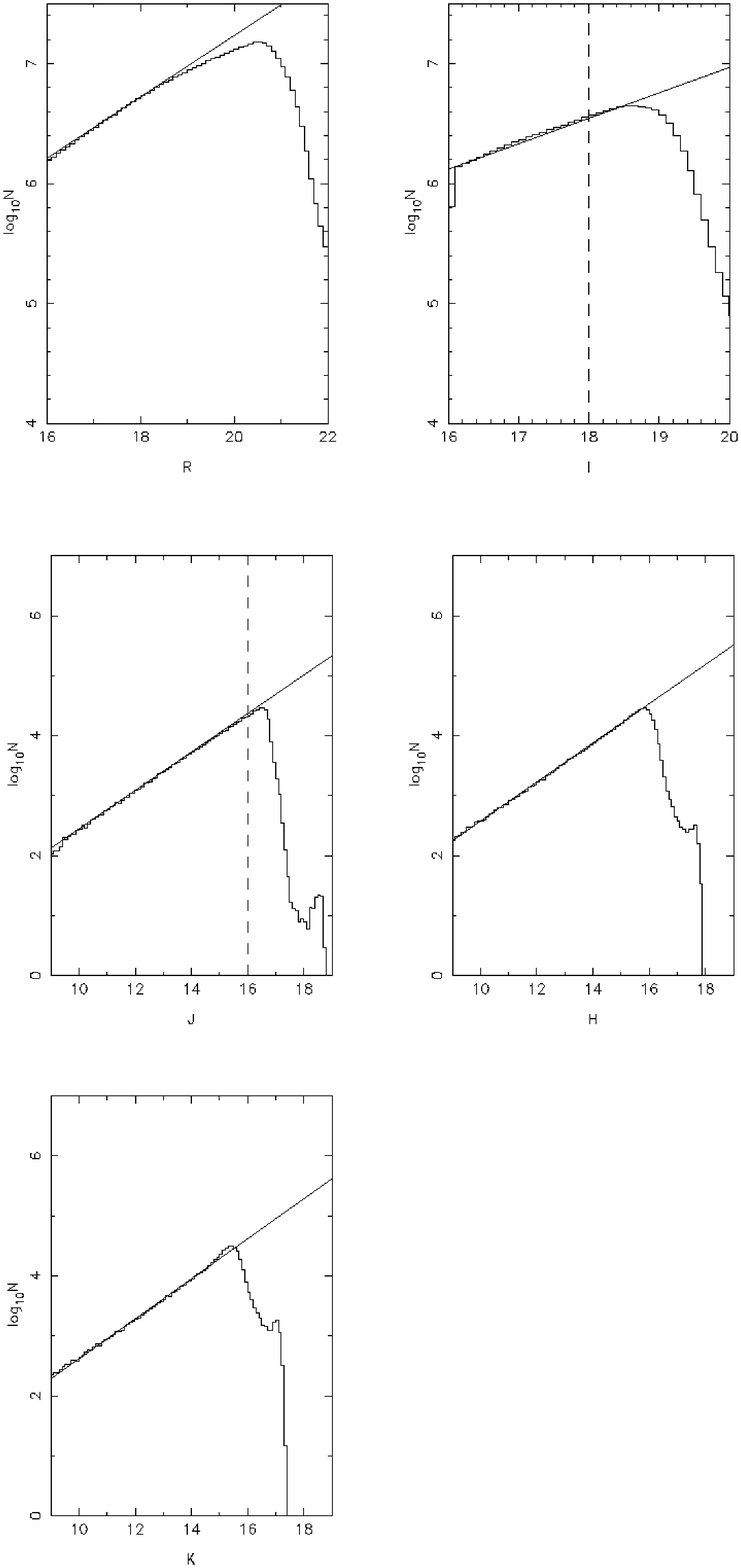}}
\end{center}
\caption{The completeness estimates for the five passbands in the survey. The histogram represents the number of objects in the actual survey and the straight line is a fit to the trend in the area where the survey is complete. The drop-off from this line is the incompleteness estimate used in the survey. The two dashed lines represent the $I$ and $J$ band limits of the SIPS survey.}
         \label{comp}
\end{figure}

The photometric selection consisted of two parts, simple magnitude and colour cuts and a simulation of the gradual drop-off in detectability near the detection limits. The simple magnitude cut takes the apparent magnitude and colour selections used in the reduction of the SIPS sample and applies them to the simulated stars\footnote{In the case of the $R-I$ cut, some objects may have moved a suffient distance between the $R$ and $I$ so they are no longer paired in the SuperCOMOS software. Hence as they would not have an $R-I$ colour in our original SIPS sample and are treated accordingly.}. The gradual drop-off is more complicated. A histogram of the logarithm of the number of objects vs. magnitude is created for each passband. In each histogram a straight line in this logarthimic space (ie a power law) is fitted to the section where the survey is believed to be complete. The shortfall from this line is taken as a measure of the detection incompleteness. The number counts and fits for all five pass bands are shown in Figure~\ref{comp}. Inspection of the 2MASS colour-magnitude diagram for the 2MASS file used for the calculation suggests the scale height is not a significant contribution to this drop-off. The detection incompleteness is then used as a probability that a simulated star will not be detected in a particular passband. Simulated stars are then said to be detected or not detected in a particular passband based on this probability. A non-detection in the $R$ band will not exclude an object as there is no requirement for an $R$ detection in the original survey. It will however affect the $R-I$ colour cut, objects with no $R-I$ colour automatically pass this cut. A non detection in any other passbands will exclude the star from the sample.

The astrometric selection fell into two categories, selection  by proper motion and selection based on sky position. The proper motion selection seems simple enough, we just selected on the proper motion of the object after a positional offset had been converted into a proper motion offset using the epoch difference for the particular sky area the object fell into. For selection based on sky position we not only included the survey area (roughly 20,000 sq. deg.) (based both on the area covered by the UKST plates and the $|b|>15^\circ$ Galactic latitude cut) but on area excluded by crowding. The crowding estimate (including the area obscured by bright stars) used the calculations from Deacon, Hambly and Cooke (2005). This calculates a probability an object in a particular area will be obscured by a bright star or by crowding. This probability is then used to include or exclude simulated stars from the selected sample. 
\subsection{Deriving underlying parameters}
While simulating the results of surveys is an interesting task for predicting the potential results the real goal is to establish what surveys actually tell us. As stated earlier, the birthrate and the mass function will affect the results of a survey. By varying these parameters and then comparing them to data we can constrain these underlying distributions. 

Say we take a mass function with a particular value of $\alpha$ and we define a birthrate parameter $\beta$ such that,
 \begin{equation} 
b(t) \propto e^{- \beta t}
\end{equation}
We then produce a series of simulations over a range of $\alpha$ and $\beta$ (steps of 0.08 in $\alpha$ between -2.0 and 2.0 and 0.008 in $\beta$ between -0.2 and 0.2.) values. We then compare these to our data using simple $\chi^2$ calculations and a range of space densities in the region 0.09-0.1$M_\odot$ (this anchoring mass range is chosen to be the same as that in Burgasser, 2004) which we will call $\gamma$. This produces a datacube of $\chi^2$ values. The values in the datacube can then be converted into a probability surface over $\alpha$ and $\beta$ by converting the $\chi^2$ values to probabilities (yielding a probability datacube) and marginalising over $\gamma$ and normalising to the total probability in the grid,
\begin{equation}
p(\alpha,\beta)= \frac{\int e^{-\chi^2/2}d\gamma}{\int \int \int
  e^{-\chi^2/2}d\gamma d\alpha d \beta}.
\end{equation}
Hence we can now marginalise further to gain information solely on $\alpha$ and $\beta$. Additionally if we know the density normalisation of the original simulations we can extract information on the density of stars by analysing the probability distribution in $\gamma$ (simply derived by marginalising the probability datacube over $\alpha$ and $\beta$). 
\section{Results}
\begin{figure}[htb]     
        \begin{center}
\resizebox{\hsize}{!}{\includegraphics[scale=1.0]{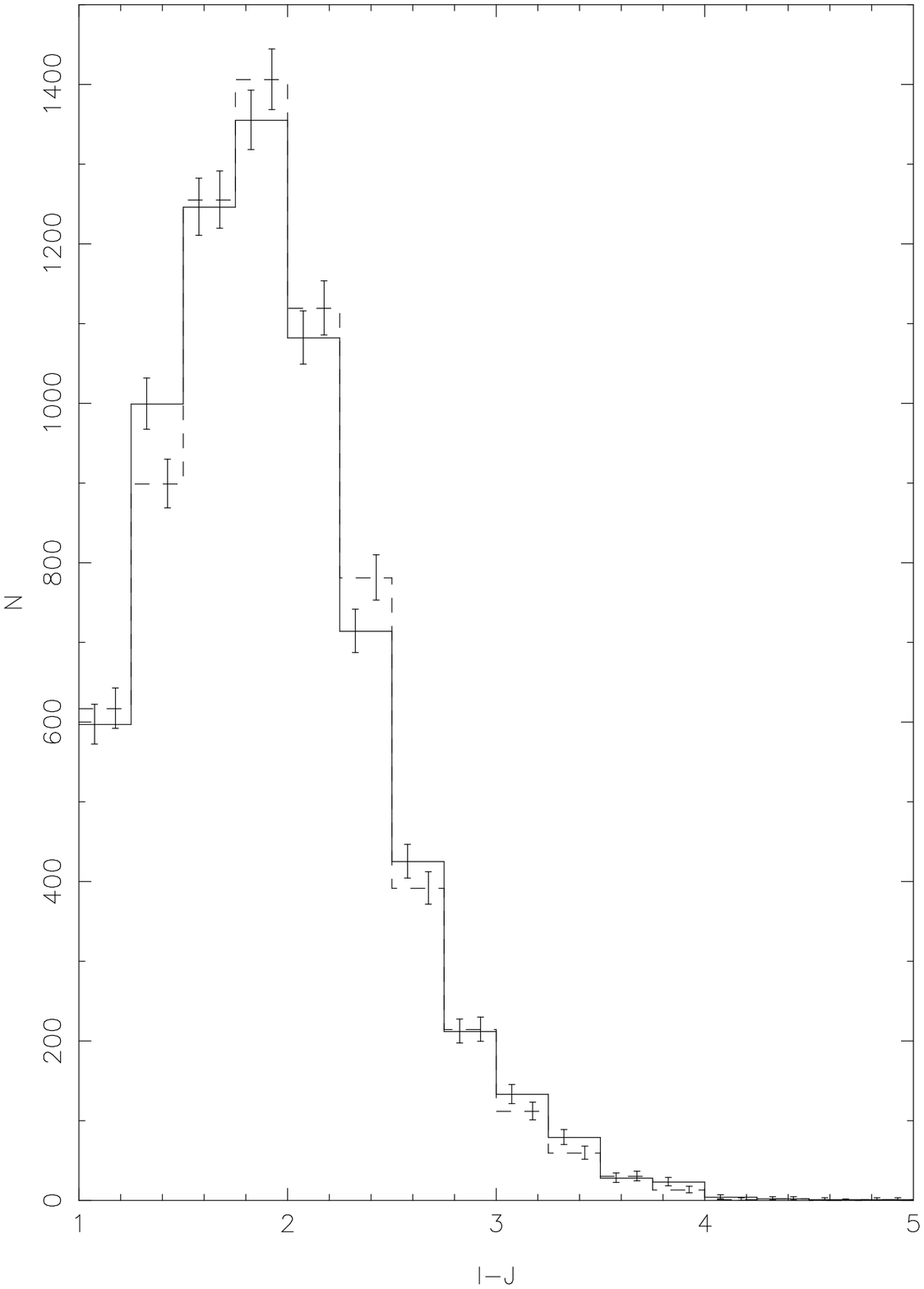}}
\end{center}
\caption{A simulation (dashed line) using the best fit parameters is shown along with the actual observed data (solid line).}
         \label{finalhist}
\end{figure}
\begin{figure*}[htb]     
        \begin{center}
\resizebox{\hsize}{!}{\includegraphics[scale=1.0]{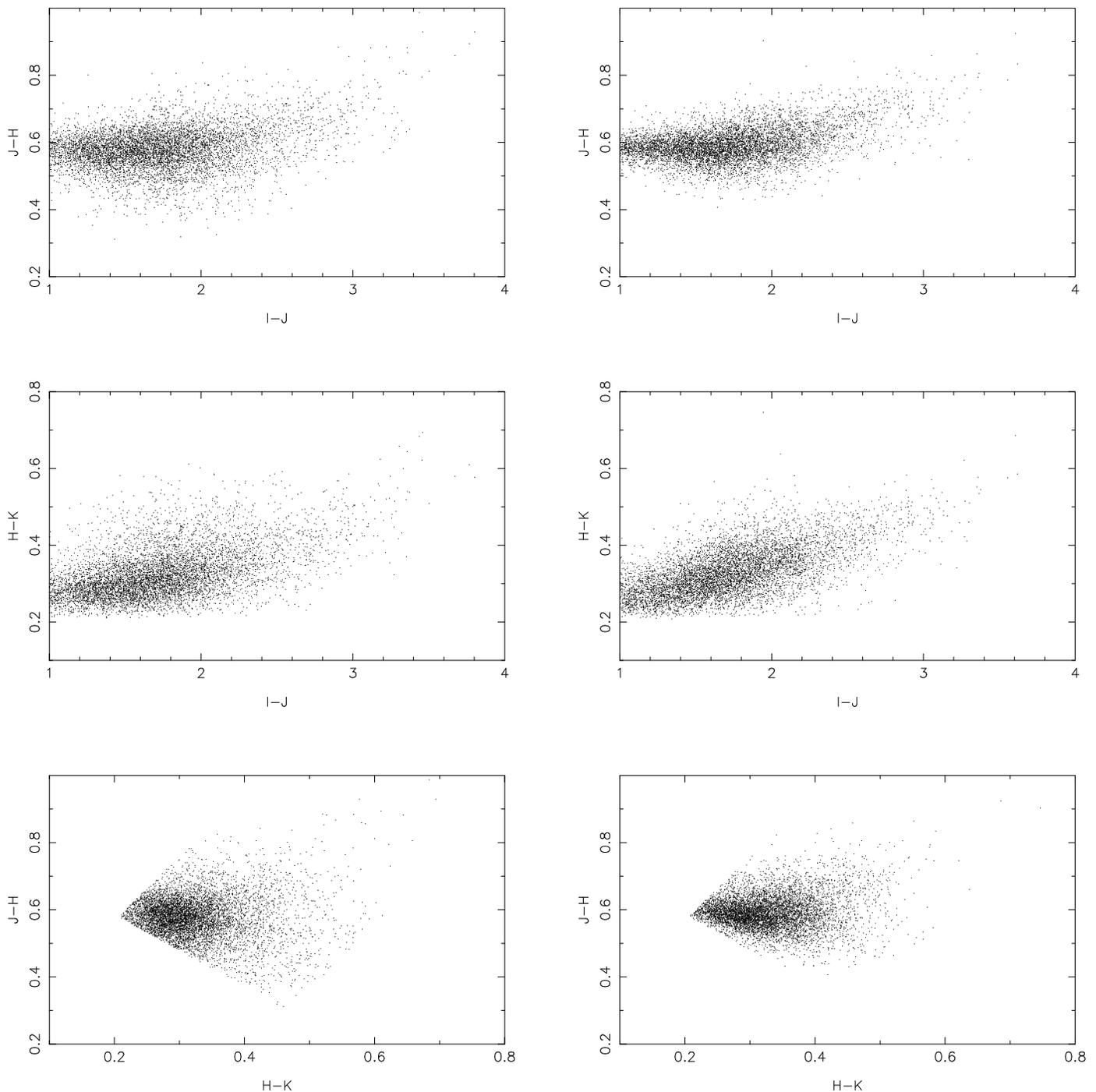}}
\end{center}
\caption{The colours of stars in the sample (left hand panels) and of simulated objects in the best fit model (right hand panels). The comparison appears good with the exception of a slightly higher scatter in $H-K$ colours in the models.}
         \label{colours}
\end{figure*}
\begin{figure*}[htb]     
        \begin{center}
\resizebox{\hsize}{!}{\includegraphics[scale=1.0]{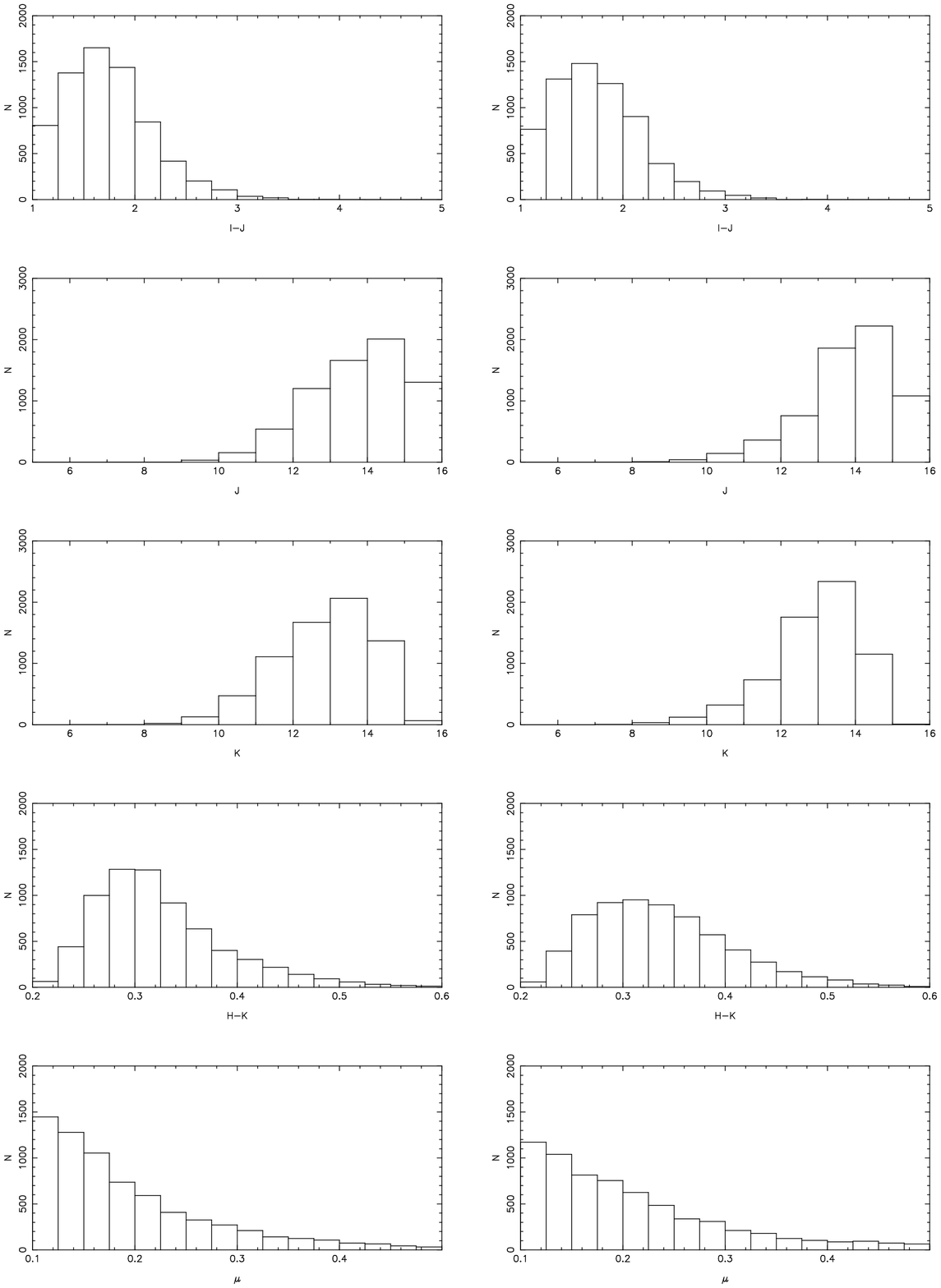}}
\end{center}
\caption{Histograms for the data (left hand side) and simulation results (right hand side). Note the slightly different brightness distributions and the higher spread in $H-K_s$ colours in the simulations.}
         \label{hists}
\end{figure*}
The grid of simulations was produced as detailed above with the survey results being represented by an $I-J$ histogram. This was originally chosen as $I-J$ colour increases into the T dwarf regime unlike other IR colours. The grid of histograms was then compared to the observed histogram to produce a datacube of $\chi^2$ values and hence probabilities. This allowed us to produce constraints on the parameters $\alpha$, $\beta$ and $\gamma$. The derived value for $\alpha$ is $-0.87\pm0.06$ and the value for the space density of stars in the range 0.09-0.1$M_\odot$ ($\gamma$) is found to be $0.0024 \pm 0.0009 pc^{-3}$. Clearly we cannot claim an accuracy on the value of $\alpha$ which is smaller than our grid's resolution (step size). Hence the value of the error in $\alpha$ must be set to 0.08. The probability distribution for $\beta$ is too noisy to produce any sensible constraint. This is most likely due to the low number of L dwarfs in the SIPS-II sample (fourteen in total) as the characteristics of the mainly Hydrogen burning M dwarfs which dominate this sample do not change rapidly over time, unlike the mostly substellar L dwarfs. The results for $\alpha$ and $\gamma$ assume no errors in the model beyond simple counting errors and do not include binarity. Hence they should be assumed to apply to the system mass function (a mass function with calculated from a luminosity function with no binary correction applies, see Chabrier, 2002). The best fit has a $\chi^2$ per degree of freedom of 1.64 and is shown in Figure~\ref{finalhist}. Additionally various colours of the individual objects in the SIPS-II sample and in the best fit simulation are plotted in Figure~\ref{colours} while histograms for various colours, apparent magnitudes and the proper motion are shown in Figure~\ref{hists}.
\subsection{Potential photometric errors in the model}
Clearly we cannot assume that our model is perfect. The most likely source of potential errors is the photometric model. To attempt to quantify this we examined the effects of a simple offset in the $I-J$ colour. We did this for two reasons, firstly $I-J$ is the colour used in the histogram for comparing results to surveys and secondly the $I$ magnitude fit has much more scatter than the $J$ or $H$ fits (see Figure~\ref{Tfit1}). Note we do not examine the $R$ band fit as, while it is as noisy as the $I$ band fit, it is not used as a cut for all stars (as some will not have a paired $R$ magnitude due to photometric incompleteness and stars moving beyond the SuperCOSMOS pairing radius of 6 arcseconds). 

A series of simulations with the best fit value of $\alpha$ and a range of $I-J$ offset between -0.1 and 0.1 magnitudes were produced. These were then compared with the simulated grid and the probability distributions in $\alpha$ produced by this process were added. The standard deviation of this combined distribution was then measured to determine the error due to the offsets. This was found to be 0.21. Hence we find that our value for $\alpha$ ($\alpha_{sys}$)for the system mass function to be $0.87\pm0.22$. Additionally our density normalisation parameter $\gamma$ has an error of 0.0008 from potential photometric offsets.
\subsection{Correction for binarity}
\begin{figure}[htb]     
        \begin{center}
\resizebox{\hsize}{!}{\includegraphics[scale=1.0]{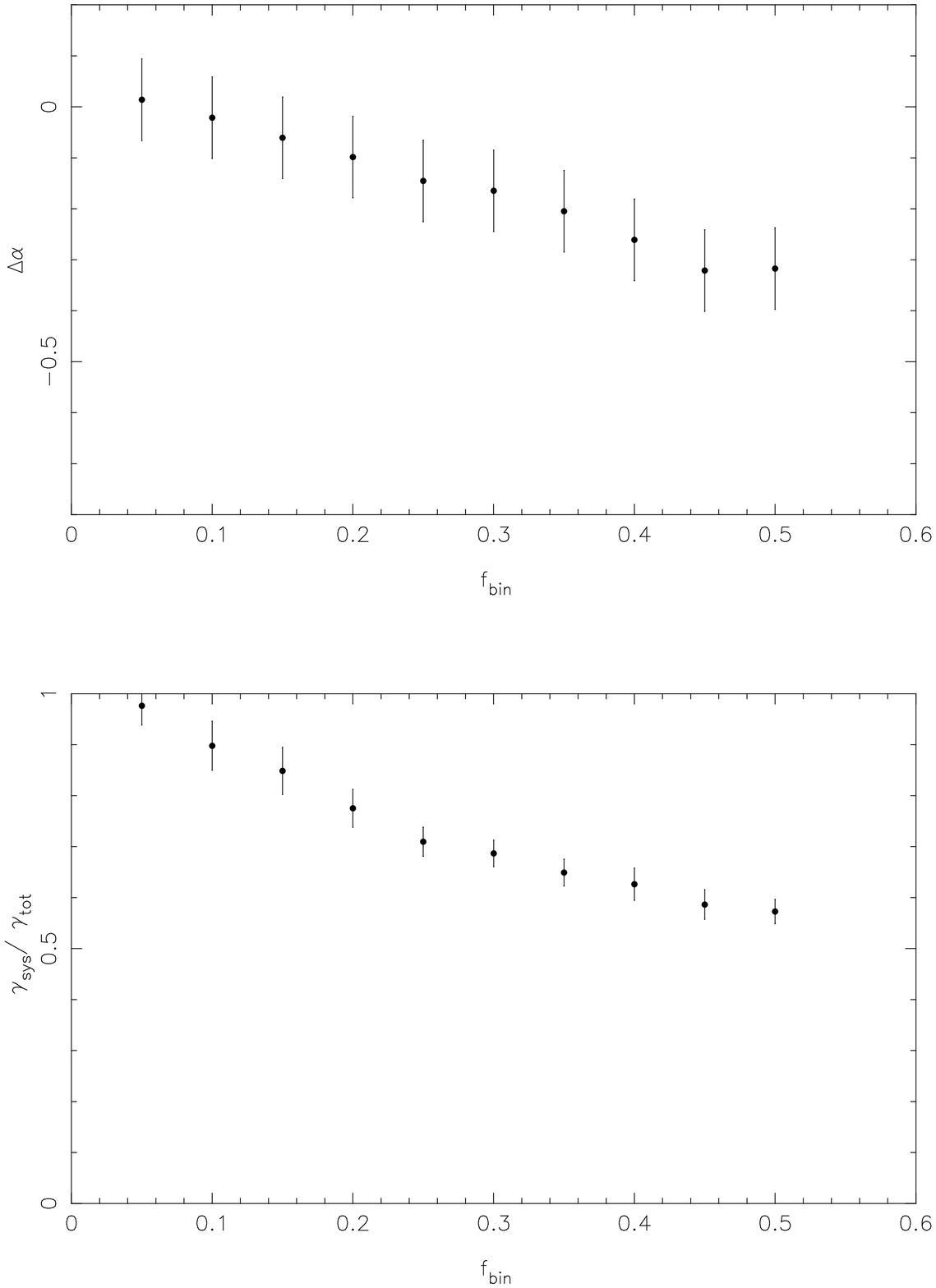}}
\end{center}
\caption{The effects on the mass function exponent $\alpha$ and the normalisation factor $\gamma$ from the inclusion of different binary fractions ($f_{bin}$) in the simulations. Note the values of $\delta \alpha$>0, this is due to noise in the simulations.}
         \label{alphabin}
\end{figure}
As stated above the mass function is strongly affected by the inclusion of unresolved binary systems. In order to remove this effect and to find the form of the individual object mass function we undertook a series of simulations. In each a proportion of the sample (the binary fraction $f_{bin}$) was assigned an unresolved binary companion. These had their masses drawn from the same mass function as the primary objects. In each passband the luminosities of the objects were added and then converted to magnitudes. This allows us to include both the brightening due to the luminosity of the unresolved companion and the effect on the colour of the unresolved object. In each of these simulations both primary and secondary objects counted towards our density normalisation factor $\gamma$. After initial test runs a value of $\alpha=-0.5$ was selected to give a result after the effects of binarity to the measured value of $\alpha=-0.85$ for the system mass function.

Once these simulations were carried out they were compared to the grid of simulations with zero binarity. The value of alpha measured then had the input value of $\alpha$ subtracted to yield the change in $\alpha$ which we shall call $\Delta \alpha$. Also the correction factor for the normalisation factor $\gamma$ was also calculated for each different value of $f_{bin}$. Figure~\ref{alphabin} shows the effect on both parameters for a range of values of $f_{bin}$. Note that we assume a value of $f_{bin}$ that does not change with distance. Clearly this is not correct as nearby objects will be more likely to be resolved. As the lower mass objects in our sample tend to be nearer by this may introduce a bias into the measurement.

In order to make the correction to yield the individual object mass function we take the binarity estimate of Maxted \& Jeffries (2005). As our survey uses 2MASS data and photographic plates which have fairly low resolution we shall assume that all binaries are unresolved. The Maxted \& Jeffries (2005) estimate of 32-45\% varies due to different underlying distributions such as the distribution of separations. For our correction we shall take the middle of this range to be our binary estimate and use the extremities of it as the one $\sigma$ error. Hence we use $f_{bin}=0.385\pm0.065$. Interpolating between points we get a value for $\Delta \alpha$ of $-0.25\pm0.1$. The uncertainty in the value of $f_{bin}$ adds an additional error of 0.09 to the determination of $\Delta \alpha$. Hence $\Delta \alpha=-0.25\pm0.13$. Applying this correction to the system value of $\alpha$ ($\alpha_{sys}$) we find the value of the exponent for the individual object mass function ($\alpha_{ind}$) to be $-0.62\pm0.26$.

Using the assumed value of $f_{bin}$ above we can also find the density of individual objects with masses between 0.1 and 0.09 solar masses ($\gamma_{ind}$). We find that $\frac{\gamma_{sys}}{\gamma_{ind}}=0.63\pm0.06$. This gives $\gamma_{ind}=0.0038\pm0.0013pc^{-3}$.
\subsection{The region of mass function applicability}
\begin{figure}[htb]     
        \begin{center}
\resizebox{\hsize}{!}{\includegraphics[scale=1.0]{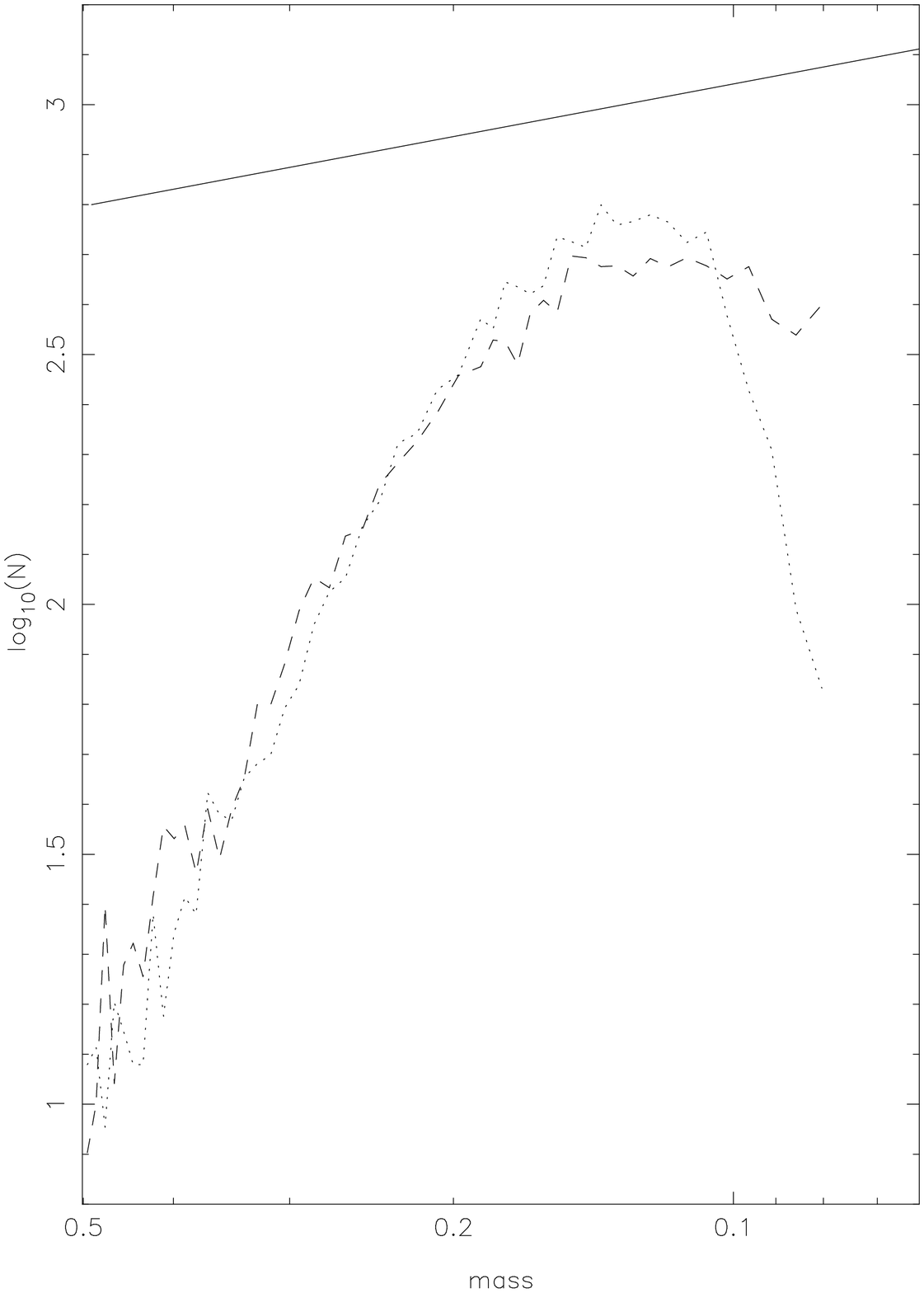}}
\end{center}
\caption{A plot showing the number of objects of different masses passing through the survey selection mechanism in the best fit simulations. The dashed line represents the best fit model after binary correction and the dotted line the best fit model with no binary correction. The solid line shows the best fit mass spectrum (a power law with an index (-1-$\alpha$). The change in shape between the mass spectrum and the range of detected masses represents the incompleteness.}
         \label{massplot}
\end{figure}
Now we have an estimate for the value of $\alpha$ we must estimate over what range of masses it is valid. As we have very few L dwarfs we cannot claim we are probing the substellar regime. Hence a lower limit of $0.075M_\odot$ seems sensible. As for an upper limit, excluding photometric errors the $I-J$ histogram cuts out at around 1.5. This equates to $M_K \sim 7.5$. Using the mass-luminosity relations from Delfosse et al (2000) this gives us an upper limit of $0.2M_\odot$. Additionally we can use our simulations to estimate the region of mass function applicability. Examining the range of masses passing through the simulation (see Figure~\ref{massplot}) we find a sharp drop below the hydrogen burning limit, a peak at around 0.1 solar masses followed by a steady decline towards higher masses.
\section{Discussion}
\begin{figure}[htb]     
        \begin{center}
\resizebox{\hsize}{!}{\includegraphics[scale=1.0]{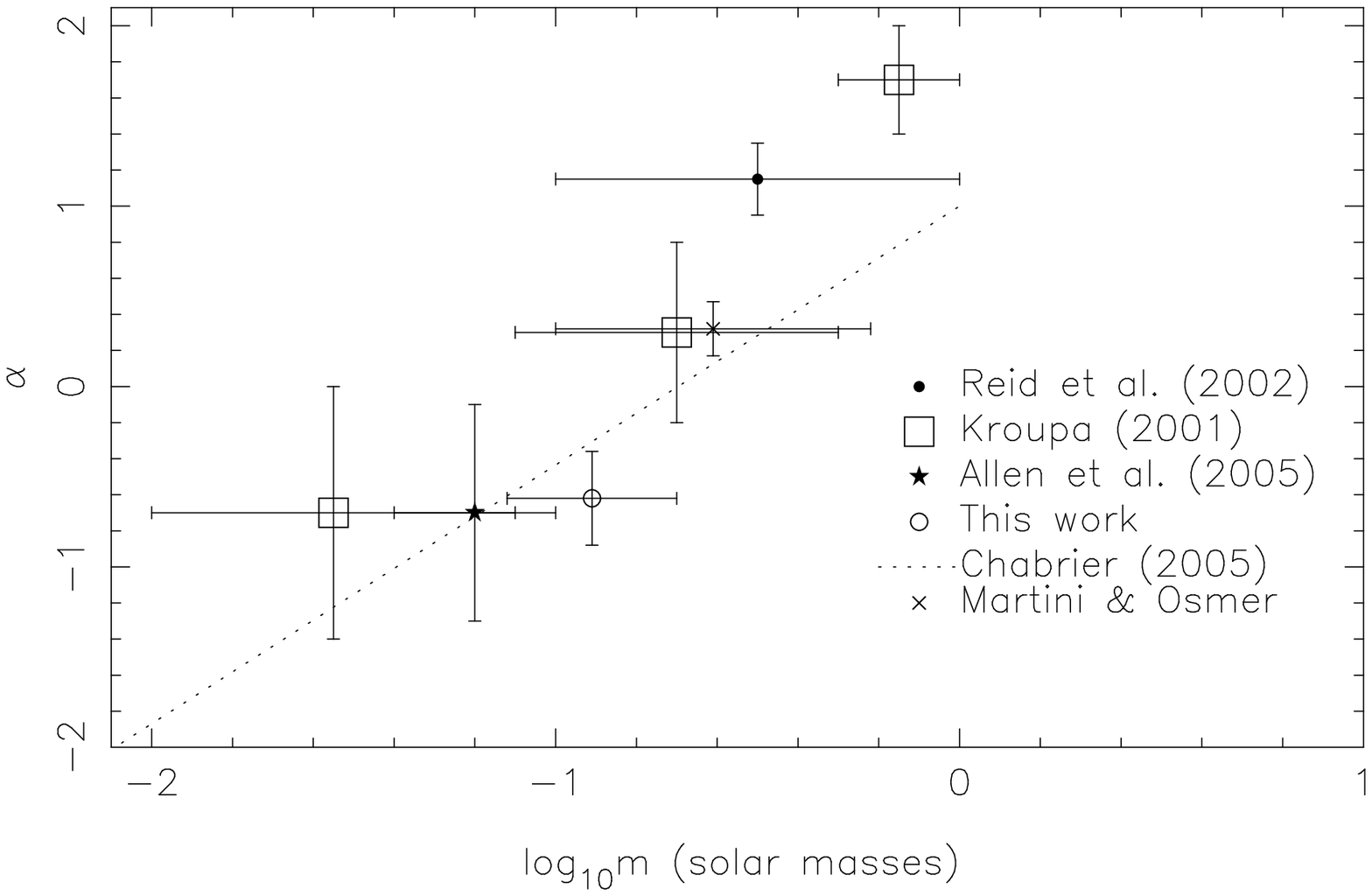}}
\end{center}
\caption{A plot showing the different values of $\alpha$ found by this and other studies. The vertical error bars represent errors in $\alpha$ while the horizontal error bars represent the range of masses over which the value of $\alpha$ is valid.}
         \label{alphaplot}
\end{figure}
In order to compare our calculated value of $\alpha$ with those of other studies an $\alpha$ plot was produced. This is shown in Figure~\ref{alphaplot}. This value of $\alpha$ differs just outside the quoted errors from those of Kroupa (2001) at the high end and middle of our mass range. However it agrees with Allen et al. (2005) at the low end and is in good agreement with Zheng et al. (2001). In the middle of the mass range the gradient of the Chabrier (2005) lognormal form differs from ours just outside the error bounds. However assuming that the errors on his parameters are of similar magnitude to those in Chabrier (2001) the two determinations agree within one sigma. We differ significantly from the studies of Reid et al. (2002) and Martini \& Osmer, these studies however cover a much larger (and mostly higher) mass range than our sample.

Burgasser (2004) uses the studies of Reid et al. (2002) and Chabrier (2001) to derive a value for the number density in the region 0.1 to 0.09 solar masses. He estimates this to be 0.0055$\pm$0.0018. Our estimate of 0.0038$\pm$0.0013 agrees with this within one sigma. The mass function calculated by Chabrier (2005) gives a value for the number density in this region of 0.0036. While no errors are quoted on the parameters of this mass function it is clear that this value agrees well with that calculated in this work.
\section{Conclusions}
We have used simulations of the low mass star population to attempt to constrain the birthrate and mass function. Unfortunately no clear constraint could be set on the birthrate. However after a correction for binarity and taking into account the potential errors in our model we found a value of -0.62$\pm$0.26 for the exponent of the mass function power law ($\alpha$). Additionally we find a constraint on the number density of stars with masses in the region 0.1-0.09 solar masses of 0.0038$\pm$0.0013. Both these results are consistent with some studies in the field.

The obvious next step for such work is to extend it to cover other surveys for low mass stars and brown dwarfs such as the UKIDSS Large Area Survey (Lawrence et al., 2006). This survey has substantially more accurate photometry compared to the UKST $I$ data used in this study. Additionally the scope of the simulations could be extended to include different stellar populations such as low metallicity halo objects, allowing a more accurate model of the local low mass population.
\begin{acknowledgements} N.R.D. is supported by NWO-VIDI grant 639.042.201 to P. J. Groot, G.N. is supported by NWO-VENI grant 639.041.405; The authors would like to thank Sue Tritton and Mike Read for their
help in selecting plates, Harvey MacGillivray and Eve Thomson for
their scanning of the material on SuperCOSMOS and David Bacon, Andy Taylor, Thomas Kitching, Hugh Jones and John Cooke for their helpful discussions. 
This publication makes use of data products from the Two Micron All Sky Survey, which is a joint project of the University of
Massachusetts and the Infrared Processing and Analysis Center/California Institute of Technology, funded by the National
Aeronautics and Space Administration and the National Science Foundation. SuperCOSMOS was funded by a grant from the
UK Particle Physics and Astronomy Research Council. This publication makes use of the SLALIB positional astronomy library (Wallace, 1998).     
\end{acknowledgements}

\end{document}